# Structure of Gel Phase DPPC Determined by X-ray Diffraction

John F Nagle[1], Pierre Cognet[1], Fernando G. Dupuy[2], and Stephanie Tristram-Nagle[1]

[1]Department of Physics, Carnegie Mellon University, Pittsburgh, PA USA 15213
[2]Instituto Superior de Investigaciones Biológicas (INSIBIO), CONICET-UNT and Instituto de Química Biológica "Dr Bernabé Bloj", Facultad de Bioquímica, Química y Farmacia, UNT. Chacabuco 461, San Miguel de Tucumán, Argentina.

10/2/18  V7

Abstract:

High resolution low angle x-ray data are reported for the gel phase of DPPC lipid bilayers, extending the previous q range of 1.0 Å$^{-1}$ to 1.3 Å$^{-1}$, and employing a new technique to obtain more accurate intensities and form factors |F(q)| for the highest orders of diffraction.  Combined with previous wide angle x-ray and volumetric data, a space filling model is employed to obtain gel phase structure at a mesoscopic level.  A new conclusion from this analysis is that the hydrocarbon chains from opposing monolayers are mini-interdigitated; this would help explain the previously well-established result that the opposing monolayers are strongly coupled with respect to their chain tilt directions. Even more detailed structural features are described that have not been obtained from experiment but that could, in principle, be obtained from simulations that would first be validated by agreement with the wide angle and the new low angle |F(q)| x-ray data.

## I. Introduction

Bilayers of the DPPC lipid have been the foremost studied proto biomembrane system. It is a benchmark against which other bilayers and membranes have been compared. The gel phase of DPPC has been especially well studied [1-7] because so much experimental information can be obtained for it compared to the fluid phase of DPPC or any other lipid bilayer. It is therefore an especially rigorous testing ground for simulations and force field development.[8-17]

Although gel phase DPPC bilayer structure has previously been reported from this lab,[4-7] we recently developed an improved x-ray diffraction method that obtains higher resolution (more orders of diffraction) than previously. This provides form factors (Fourier transforms of the electron density profile) which are primary x-ray data that simulators should compare to, and sometimes have.[10, 14, 18] After presenting the new form factors, we combine them with wide angle x-ray and volume data to obtain structural models.

## II. Materials and Methods

### A. Sample preparation, sample chamber and x-ray sources

DPPC (1,2-dipalmitoyl-*sn*-glycero-phosphocholine) was purchased from Avanti Polar Lipids (Alabaster, AL) in the lyophilized form and used as received. Organic solvents were high-performance liquid chromatography grade from Aldrich (Milwaukee, WI).

Oriented samples consisting of stacks of approximately 1600 bilayers were prepared using the "rock and roll" method.[6, 19] Four mg of lipid in organic solvent, chloroform:methanol (2.5:1, v/v) or trifluoroethanol:chloroform (2:1, v/v), was deposited onto a Si wafer (15 mm by 30 mm). The wafer was heated at 40 °C and maintained in a warm atmosphere inside a glove-box, while rocking the substrate. After rapid evaporation while rocking the substrate, an immobile film formed which was then further dried for several days to evaporate residual organic solvent. The samples were trimmed to occupy 5 mm by 30 mm within the middle of the Si substrate. The thickness of the sample (used for the x-ray absorption correction[20]) was estimated by calculation from the lipid mass and substrate area covered and the amount of water required to obtain the measured lamellar repeat D spacing. The sample was mounted in our x-ray sample chamber which provides greater than 99% relative humidity. Even greater RH was obtained with a Peltier element underneath the Si wafer which, by cooling the sample relative to the vapor, allowed tuning the D spacing up to the same fully hydrated lamellar repeat spacing D as multi-lamellar vesicles (MLV) immersed in water.[20, 21] Such tuning is quite delicate; typically, D spacing varied somewhat during the course of taking many exposures. Data were taken in the D range of 60.8 - 62.3 Å for the most recent oriented sample which provided the most important share of data used in this report. For these data the x-ray source was G1 line at the Cornell High Energy Synchrotron Source (CHESS). A W/$B_4$C multilayer monochromator selected wavelength 1.096 Å with a spread of 1%. Earlier studied oriented samples were prepared in much the same manner and irradiated with Cu$K_\alpha$ x-rays; these data were consistent with the higher quality data reported in this study.

Unoriented multi-lamellar vesicles (MLV) in excess water (3:1) were drawn into 1 mm diameter thin-walled glass capillaries. The fully hydrated D spacings were 63.2-63.6 Å. A sample mixed with a small concentration of the dehydrating agent polyvinylpyrrolidone had D = 60.9 Å. The x-ray source for two samples studied some years ago were from a fixed tube Rigaku source as described earlier[20] and were from a Rigaku RUH3R rotating anode for three later samples, including a recent one for this study. The x-rays for all the MLV samples were Cu $K_\alpha$ radiation with wavelength 1.5418 Å.

The temperature was 20°C for all data in this paper.

B. Data acquisition

Two different methods were employed for oriented samples. The first method has been employed in this laboratory for some time to collect diffuse scattering intensity for fluid phase samples.[21] Gel phase DPPC has very little diffuse scattering intensity, but this method works equally well to obtain the intensity of the lamellar orders of diffraction that are the traditional data obtained from so-called small angle x-ray diffraction (usually called SAXS, which we have preferred to call LAXS for Low Angle because we obtain peaks at larger angles than the usual SAXS regime; in this study the angle of the highest

LAXS order is about as large as the wide angle WAXS scattering from the hydrocarbon chains). The flat oriented sample was rotated about an axis that is perpendicular to the beam, parallel to the Si wafer, and located within the sample.[21] The lower limit of the rotation was -1.6°, at which the substrate completely blocks the sample from the x-ray beam; the upper limit was 11°, corresponding to a not detected h=21$^{st}$ order. An area detector ("Flicam", Finger Lakes Instrumentation, Lima, NY) collected x-ray intensity as the rotation motor ran back and forth between the two limits at a nominal maximum speed which was calibrated to ensure a complete cycle every 1.5s so that exposure times of 1.5n seconds, integer n, would sweep through all angles in the range 2n times. The limits of rotation were chosen widely so that the motor speed, which had to slow down to reverse near the limits, was uniform over pertinent angles from 0 to the highest observable order (h=13, $\theta_h \approx 6.7°$); the negative angle at which the rotation slowed down had no lamellar scattering and there was negligible scattering due to mosaically misaligned domains in the h<15 range near the maximum rotation angle. Fig. 1 shows scattering obtained by this method.

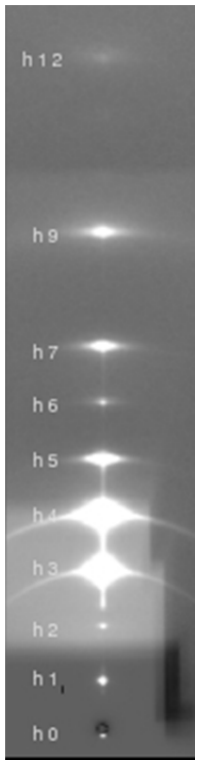

Fig. 1. Background subtracted diffraction peaks from DPPC gel phase at T=20 °C using continuous rotation of the sample. Repeat spacing D was 60.8 Å. The first order h1 and the second order h2 peaks were attenuated by a factor of 625 by 100 μm Mo. The beam at h0 was attenuated by a factor of 2 million. Background was obtained from a fixed negative angle exposure. The h=10 and h=11 intensities were quantifiably non-zero but were too small to be seen at this gray scale. Mosaic spread is evident for the overexposed h=3 and h=4 orders.

We have also used a new method for oriented samples in this paper. This method takes advantage of the fact that oriented samples are never perfect, but have mosaicity consisting of many misoriented microdomains within the footprint of the beam. The angular distribution of this apparently continuous mosaicity is closer to Lorentzian than to Gaussian,[22] so an exposure at a fixed substrate angle $\theta_F$ not equal to a Bragg angle $\theta_h$ still shows peaks for the $h^{th}$ order with the relative intensity decreasing gradually for those peaks with Bragg angles further from the fixed exposure angle, i.e., for larger $|\theta_F - \theta_h|$. The new method sets the fixed angle midway between the Bragg angle of two orders, $\theta_F = \frac{1}{2}(\theta_{h1} + \theta_{h2})$. Then, the ratio of the intensities of those two orders is identical to what would be obtained by the first method above because the mosaic distribution is symmetrical. An advantage of this second method is that the actual intensity $I_h$ of an order with $\theta_h$ close to $\theta_F$ is greater than for the rotation method. Most of the fixed angles were chosen midway between two adjacent orders $\theta_F = \frac{1}{2}(\theta_h + \theta_{h+1})$ to obtain the intensity ratio $I_{h+1}/I_h$. Figure 2a shows the result from which we obtained the ratio $I_{11}/I_{10}$ and also the ratio $I_{12}/I_9$ from the much stronger intensities of the h=9 and h=12 peaks. (To obtain $I_{10}/I_9$, $\theta_F$ was set to $\frac{1}{2}(\theta_{10} + \theta_{11})$ – not shown.) Figure 2b shows the result when the fixed angle was chosen to be $\theta_F = \frac{1}{2}(\theta_7 + \theta_9)$ to obtain the $I_9/I_7$ ratio. This shows that the h=8 order was extinct because the only scattering has the shape of the beam and is clearly weak specular scattering from the substrate.

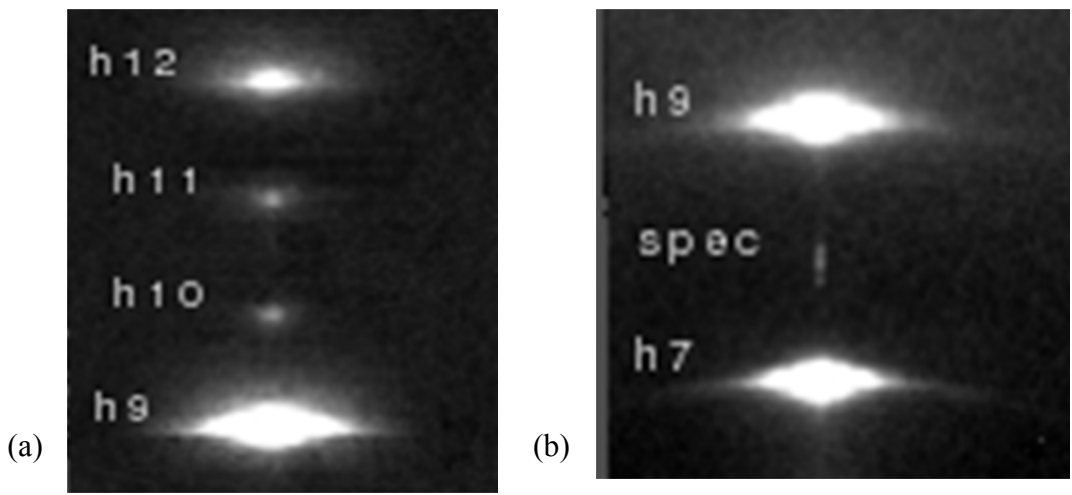

Fig. 2. Examples of scattering at fixed angles to obtain ratios of intensities of pairs of orders. (a) $I_{12}/I_9 = 0.23$ and $I_{11}/I_{10} = 0.65$. and (b) $I_9/I_7 = 1$, $I_8 = 0$. D = 60.8 Å.

For both methods for oriented samples, longer exposures could be taken for the weaker higher orders by adjusting a motorized x-ray absorber to attenuate the much stronger lower orders that would otherwise have saturated the area detector. Also, the beam size was 0.2 mm wide (in the plane of the stack of bilayers) by 1 mm high (perpendicular to the substrate for zero rotation angle); this guaranteed that the 5 mm part of the sample in the direction of the beam remained in the beam for all rotation angles, thereby assuring that the appropriate Lorentz factor was proportional to q for the first method. Previous studies from this lab[7] have reported the average in-plane domains as large as 2900 Å and roughly 600 Å perpendicular to the substrate in the fluid phase,[23] equivalent to a domain volume of about 1 $\mu m^3$. As the sample volume in the beam is $10^7$ $\mu m^3$, one expects the effectively continuous mosaicity distribution that was observed[22].

## C. Repeat spacings and peak intensities

To obtain the lamellar D spacing from each exposure, peak positions in pixels $p_h$ were entered into an app in our visualization software for as many orders h as could be robustly located. The app used the wavelength, the sample to detector distance (S=179.3 mm) and the pixel size ($pix_s$=0.7113 μm) to calculate the values of D and the beam position $p_0$ that provides the best least squares fit to

$$p_h = p_0 + (2S/pix_s)\tan(\sin^{-1}(\lambda h/2D)) \tag{1}$$

for the entered h orders. The reciprocal space locations $q_h$ of the $h^{th}$ order was then $2\pi h/D$. One advantage of our experimental setup for oriented samples is that the D spacing could be varied by varying the relative humidity and this provided many q values for the Fourier form factor F(q).[20]

Background, from He and water vapor in our chamber and from the mylar windows and from upstream gases in the CHESS flight path, was subtracted from each peak separately using an app in our visualization program. Slowly varying additional background intensities occurred in ranges of $p_x$ to either side of a peak. Both these off-peak $p_x$ regions were fit simultaneously for each value of $p_z$ with either a constant or a second order polymonial. Interpolation across the $p_x$ range of the peak then provided the background that was subtracted from the intensity in the region of the peak. The integrated peak intensity was then summed within a rectangular pixel area; the width of the rectangle was increased with increasing order to ensure inclusion of the same number of mosaic domains for each order. Values of $I_h$ exceeded $10^6$ for h=1 and also for some of the higher orders collected for longer exposure times. The relative uncertainties from the background subtraction for each peak were typically much less than 1% for these orders.

However, repeated exposures typically yielded larger uncertainties. Standard errors $\sigma_h$ for each order h were obtained from exposures that had very nearly the same D spacing. There were typically only up to five repeated exposures due to the difficulty of maintaining the same value of D. This may have contributed to the relative errors $r_h \equiv \sigma_h/I_h$ varying considerably with h due to drifting of the relative humidity, but the overall trend with measured intensity was consistent with the theoretical expectation $r_h = c/I_h^{1/2}$. The ensuing empirical value for c of approximately 30 was then used to assign relative uncertainties, but with several exceptions. The first exception was when the peak was unobservable, such as h=8. Zero intensity is, of course, a real and important result, but the assigned uncertainty should clearly not be zero or infinity; it was instead taken to be equal to the uncertainties assigned to the well quantified h=7 and h=9 orders. Similar uncertainties were assigned to the weak, but barely measurable h=6 peak and some of the higher order peaks that had small intensities at some D spacings and that had much larger fluctuations in $r_h$. The uncertainties in the intensities of these weaker peaks were consistent with background subtraction being the major source of uncertainty.

Another exception was that much larger uncertainties were assigned to the h=1 and h=2 peaks for oriented samples than would be inferred from their very large intensities. These intensities could appear too large due to the very large specular intensity at low angles. On the other hand, diffraction from mis-oriented domains is cut off by the rotating substrate when the domains are mis-oriented by more than the Bragg angle $\theta_h \approx$ 0.5h degrees in our setup. The width of the mosaic distribution for our oriented samples is difficult to obtain accurately and is subject to some ambiguity due to long tails in the mosaic distribution,[22] but supposing that it could be as large as 0.5 degrees for this gel phase sample would substantially reduce the intensity $I_{h=1}$ but would hardly reduce $I_h$ for h≥3. We did observe some reduction in the ratio of $I_1/I_3$ by comparison with intensity ratios obtained from multilamellar vesicles (MLV) in capillaries which do not suffer from this cutoff artifact.

The MLV samples provide excellent intensities up to h=5. Uncertainties were assigned from the uncertainties in background subtraction. Overlap of the results from MLVs and oriented samples for the h=3-5 peaks provides the bridge between the much better results for the high h orders of oriented samples and the MLV results for h=1 and h=2 that were not affected by the cutoff artifact.

Fixed angle exposures give the ratios of the intensities $I(h_a)$ and $I(h_b)$ when the angle is ideally half way between the Bragg angles for peaks $h_a$ and $h_b$. However, our rotation motor only had steps of 0.05 degrees, so there was always some difference from the ideal angle; we measured this difference using the location of the specular reflection from the substrate. By varying the angle for a few fixed angle exposures and fitting to the intensity ratio, a correction formula was devised, but this correction could vary for exposures taken at different locations on the sample that had different mosaic spreads. To compensate, uncertainties were assigned using the same $r_h = c/I^{1/2}$ formula that was applied to exposures taken with continuous rotation. Although this may have overestimated the uncertainties for the fixed angle data, they are also more heavily weighted for the highest orders because their peaks were more intense than those collected with continuous rotation.

### D. Conversion of intensities I(q) to form factors F(q)

The quantities of final interest are not the intensities, but the form factors $F(q_h)$, where $q_h = 2\pi h/D$. For unoriented MLV samples one may conveniently write

$$|F(q_h)| = (q_h^2 I_e(q_h))^{1/2}/K_e \qquad (2)$$

where $q_h^2$ is the Lorentz factor in the small angle approximation and $K_e$ is the usual scale factor which takes account of experimental conditions such as total x-ray exposure and amount of sample. For oriented samples the corresponding Lorentz factor is $q_h$ instead of $q_h^2$ and there is also an absorption correction factor because x-rays that scatter at low angles with respect to the substrate travel further within the sample on average than x-rays that scatter at higher angles.[20] Performing these corrections to the intensities and using Eq. (2) gave the values of $|F_e(q_h)|$ for each exposure e, up to the undetermined scale factor $K_e$. These relative values of $|F_e(q_h)|$ are the primary data from x-ray diffraction. The final $|F_e(q_h)|K_e$ and their uncertainties propagated from the uncertainties in the intensities are given in tabular form in Supplementary Material.

### E. Structural Modeling

Modeling, required to obtain salient structural properties, used the SDP software program.[24] Like its predecessors[4, 25], the input is the intensities $I_e(q)$ and their uncertainties for the 28 used data sets, including two older published data sets[1, 4] and three previously unpublished MLV data sets from this lab. SDP performs the absorption and Lorentz corrections in the preceding paragraph to obtain initial values of $|F_e(q_h)|K_e$ and it calculates the initial model form factors $F_M(q_h,P)$ where P denotes the values of the many initial parameters in the model. The nonlinear least squares fitting program uses simplex minimization to search for the values of the model parameters and the experimental scale factors $K_e$ that minimize

$$\chi^2 = \sum_e \sum_h \sigma_{eh}^{-2} (|K_e F_{eh}(q_{eh})|^2 - |F_M(q_{eh},P)|^2)^2 + W \qquad (3)$$

where $\sigma_{eh}$ is the experimental uncertainty for each order h and exposure e. W is a Bayesian penalty term for soft constraints on the model parameters to satisfy other data in addition to the low angle intensities. The model thereby estimates the phase factors ($\pm 1$ for symmetric bilayers) for all values of q and the scale factors $K_e$ for each experimental data set.

The primary molecular model chosen for this work parsed DPPC into the following components: Gaussians with three parameters each (height, width and center) to represent (a) the carbonyl-glycerol moiety (67 electrons), (b) the phosphate (47 electrons), and (c) the choline moiety (50 electrons), (d) a Gaussian constrained to the center of the bilayer (two parameters) to represent the two terminal methyls on the hydrocarbon chains (18 electrons) and (e) a symmetric combination of error function with three parameters to represent the width, edge and height of the 28 methylenes on the hydrocarbon chains (224 electrons). This model therefore nominally has 14 spatial parameters, but there is a W term that penalizes the unphysical occurrence of negative volumes of any component at various values of z. Other constraints that contribute to W in Eq. (3) will be described in subsection III.B. Also, the volume of the lipid $V_L$ was constrained to the measured volume (1144 Å$^3$).[26] The volume of the headgroup, defined as the sum of components (a), (b) and (c), was constrained to the value $V_L$ = 331 Å$^3$ which was obtained from analysis of the wide angle x-ray (WAXS) data. The Appendix includes a table of published results for $V_H$ and other quantities derived from older WAXS data. Although new WAXS data were collected in the present study, they do not differ significantly enough from the earlier data to warrant detailed analysis.

## III. Results

### A. Form Factors

Figure 3 shows the continuous form factor obtained by using SDP to fit all the data sets simultaneously with the volumetric constraints, the number of electrons in each component, and, finally, a constraint on the hydrocarbon thickness obtained from the WAXS results that is contained in the W term in Eq. (3). Something like this particular W constraint is needed to prevent the trivial solution $\chi^2 = 0$ that consists of a featureless bilayer, F(q)=0, and all scaling factors equal to 0. It may be noted, however, that $D_C$ can be changed significantly without a significant effect on the $\chi^2$ of the fit. In any case, this fit does not give realistic values of the model parameters for the bilayer. Instead, it is a nearly free fit to all the data that allows one to determine the consistency of the different data sets. The resulting reduced value of $\chi^2$ (5.7) differs considerably from the ideal value of 1. *A priori*, this could mean that our estimated uncertainties $\sigma_{eh}$ in Eq. (3) are too small or that the constraints in the model are still too restrictive, but we are inclined to believe that it is due to inconsistencies in the 28 data sets that were fit. Note that there were 83 independent peak intensities, even after subtracting the 28 relative scale factors $K_e$; this is far more than the 14 spatial parameters in the electron density model. The visual fit of all these data to the continuous F(q) transform in Fig. 3 is excellent in this field. The largest outlier in Fig. 3 is the 10$^{th}$ order of Torbet and Wilkins.[1] Earlier unpublished work in this lab also indicated their $I_{10}$ intensity was too large by a factor of four. Although we display that value, we assign a large uncertainty to it in our fits so that it does not much affect the resulting model $F_M(q)$.

The value of the form factor at q=0 is given by[27]

$$F(0) = 2(n_L - \rho_w V_L)/A_L \qquad (3)$$

where $n_L = 406$ is the number of electrons in DPPC, $\rho_w = 0.333 e/Å^3$ is the electron density of water at T=20°C, $V_L$=1144 $Å^3$ is the experimental volume of DPPC and the area/lipid at the interface, $A_L = 47.3$ $Å^2$, has been determined by wide angle scattering as reported in the Appendix. Eq. (3) pins the sign of F(q) at q=0 and it provides the overall scale for F(q).

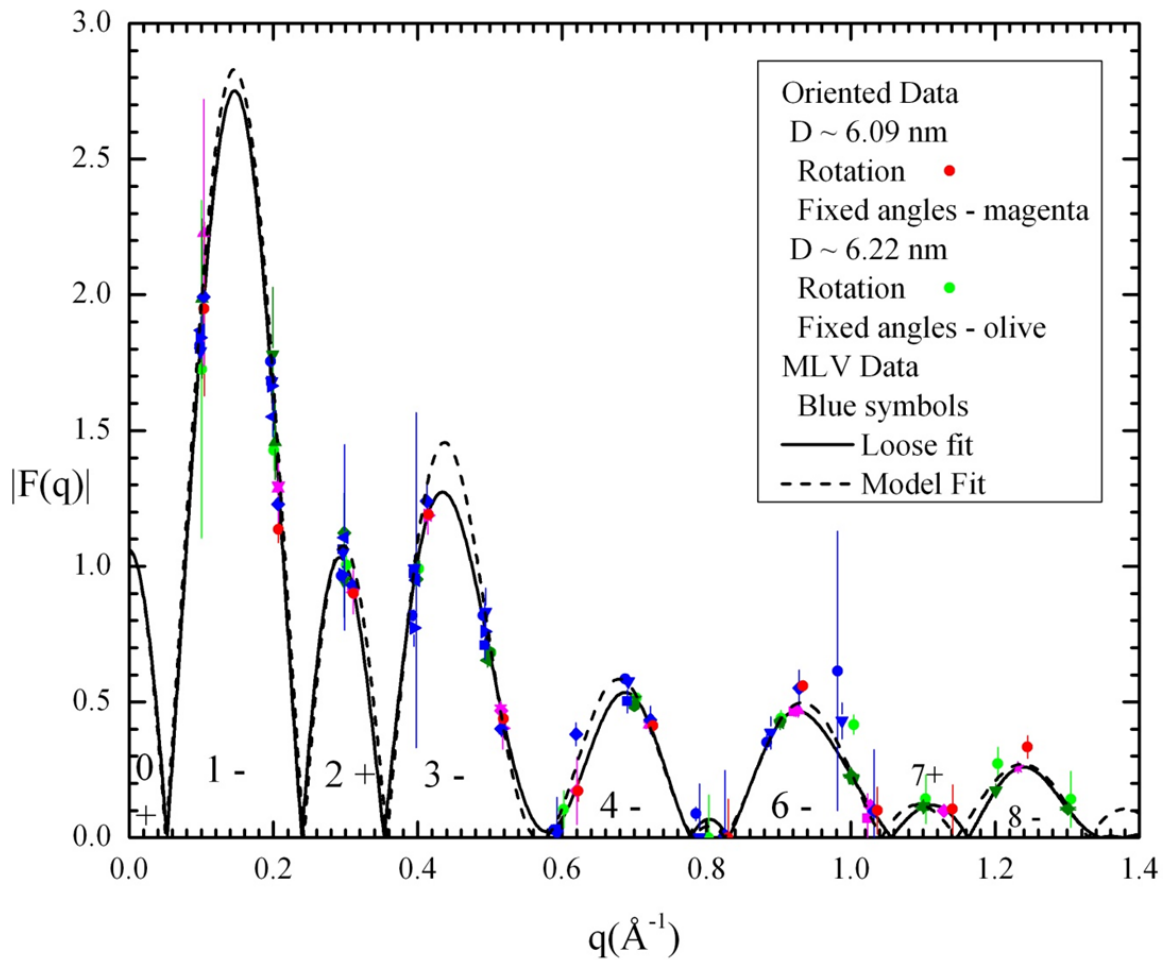

Fig. 3  Data are shown by symbols with uncertainties (vertical lines). The circles were obtained while rotating the sample for repeat spacing D=60.8 Å (red circles) and D=62.2 Å (green circles). Corresponding fixed angle data have magenta symbols and olive symbols respectively and different pairs of fixed angle data have different

shaped symbols. Blue symbols show unoriented MLV data with different shaped symbols for each data set - circles[1], squares[4], and four with other symbols from this lab. The solid curve shows the continuous $|F_M(q)|$ (electrons/Å$^2$) vs. q (Å$^{-1}$) obtained from a loosely constrained fit. The dashed curve shows $|F_M(q)|$ for a model that has realistic components. The lobes are numbered from 0 to 8 and the corresponding signs of F(q) are indicated.

As q increases from 0 in Fig. 3, F(q) decreases and becomes negative for the h=1 and h=2 orders. We will call this q region of negative F(q) the first lobe. There are then seven subsequent lobes as q increases in Fig. 3. This is the first study to report the two highest q lobes for DPPC gel phase. Note, however, that the 5$^{th}$ lobe is very small and it could disappear if a different molecular model were employed. Then there would only be two lobes, like the two between q=0.35 and 0.75 Å$^{-1}$, that would replace the 4$^{th}$-6$^{th}$ lobes in Fig. 3. Of course, such a difference in the number of lobes is only due to whether F(q) changes sign near its absolute minimal nodes and such small changes do not incur large structural differences. This suggests that one should consider the number of extrema in a plot of signed F(q); that number is 10 for both the |F(q)| lines in Fig. 3.

Fig. 3 shows the general trend that the lobes become smaller with increasing q. In addition to the usual decrease with increasing q that occurs even for crystals due to the atomic form factors, lipid bilayers immersed in water at room temperature undergo fluctuations and have intrinsic disorder, even in the gel phase. The real space average electron density profile is therefore smoothed at the shorter length scales, and that reduces F(q) at high q. This is especially evident for the F(q) of fluid (so-called liquid crystalline) bilayers for which the amplitudes of the lobes generally become negligible for q larger than about 0.8 Å$^{-1}$. In contrast, for the DPPC gel phase, Fig. 3 shows that there are quite robust lobes even for q as large as 1.24 Å$^{-1}$. This means that there is detailed structure at a smaller length scale in the DPPC gel phase than in typical fluid phases.

## B. Structure

The unconstrained fit shown in Fig. 3 does not give realistic values for structural parameters. The goal in this subsection is to obtain structural quantities such as the locations of the molecular components, the width of their distributions along the bilayer normal and their volumes. Remarkably, it turns out that it is only necessary to impose one constraint in addition to the loose fit in the previous section. The hydrocarbon boundaries with the interfacial regions are characterized by a width $\sigma_{HC}$ which takes into account both statistical disorder of lipids relative to the bilayer center and also the well known result that the sn-1 and sn-2 chains are inequivalent,[2] with the sn-1 chain embedded deeper in the bilayer so the location of the hydrocarbon interface is different for the two chains. These effects are modeled with a width $\sigma_{HC}$ defined as the interval over which the hydrocarbon volume probability increases from 0.31 to 0.69.

Figure 4 shows the electron density profile $\rho(z)$ of the bilayer. The most robust quantity from x-ray diffraction is the head-head spacing $D_{HH}$, defined as $2z_{max}$ where $\rho(z_{max})$ is the maximum value of $\rho(z)$. As expected, $D_{HH}$ is only larger by 0.2 Å in this constrained fit than in the loose fit because the two $F(q)$ are so similar. Other constraints have also been considered that give sensible, although different values for some other features, but the value of $D_{HH}$ varies by less than 0.1 Å.

Figure 4 also shows the electron density profiles of the components. The position of the phosphate component $z_{ph}$ = 22.8 Å is located farther from the bilayer center than the peak in $\rho(z)$ at $D_{HH}/2$ = 22.6 Å. The maximum of $\rho(z)$ is shifted to a smaller z value than the location of the phosphate by the addition of the carbonyl-glycerol CG component. This shift $D_{HH}/2 - z_{ph}$ is smaller than for fluid phases because the gel phase peaks are better separated because their widths are smaller, with full widths at half maximum $\sigma_{Ph}$ = 3.9 Å and $\sigma_{CG}$ = 4.8 Å.

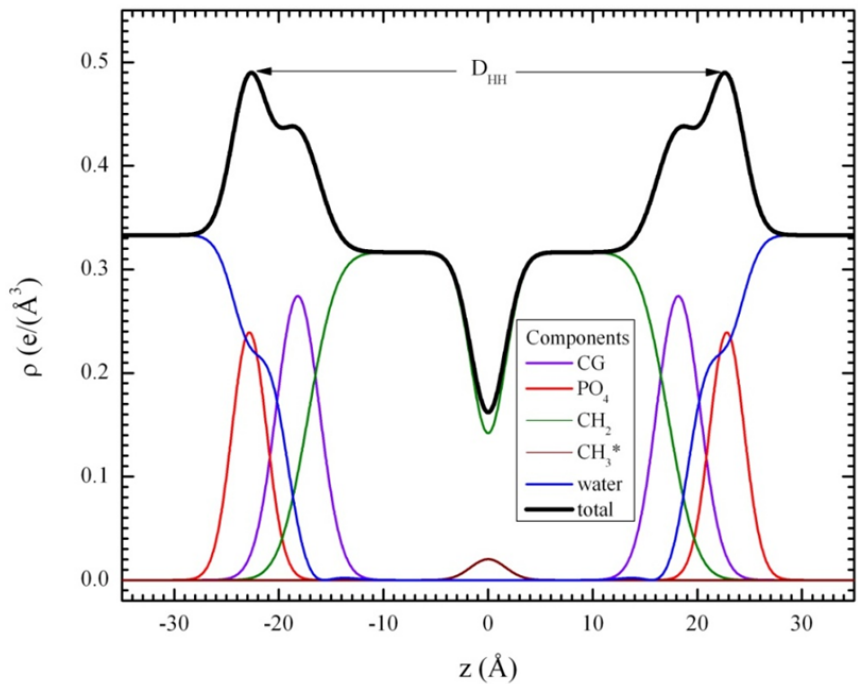

Fig. 4. Electron densities versus distance z from bilayer center.

As usual for gel phases, there is a prominent methyl trough in the total electron density $\rho(z)$ in the center of the bilayer. However, Fig. 4 employs a different visualization of the hydrocarbon components than in past publications where each chain was parsed into 14 methylenes and one terminal methyl with 9 electrons. In Fig. 4, each chain is parsed into 15 methylenes, each with 8 electrons, and a left-over methyl-like component with only 1 electron. The figure legends allude to this by adding an asterisk to the $CH_3^*$ component. The new idea is that the terminal methyl at the end of an all-trans chain looks very much like a methylene except for an excess hydrogen extending further along the chain. Importantly, this new parsing made no difference to the results for any other components. However, it suggests a new interpretation. The distinction is best viewed in Fig. 5 which shows the volume fractions of the components. At the center of the bilayer, half the volume is occupied by methylene-like components and half by the pseudo $CH_3^*$. This supports mini-interdigitation of the chains from opposing monolayers.[6] This is elaborated further in the Discussion.

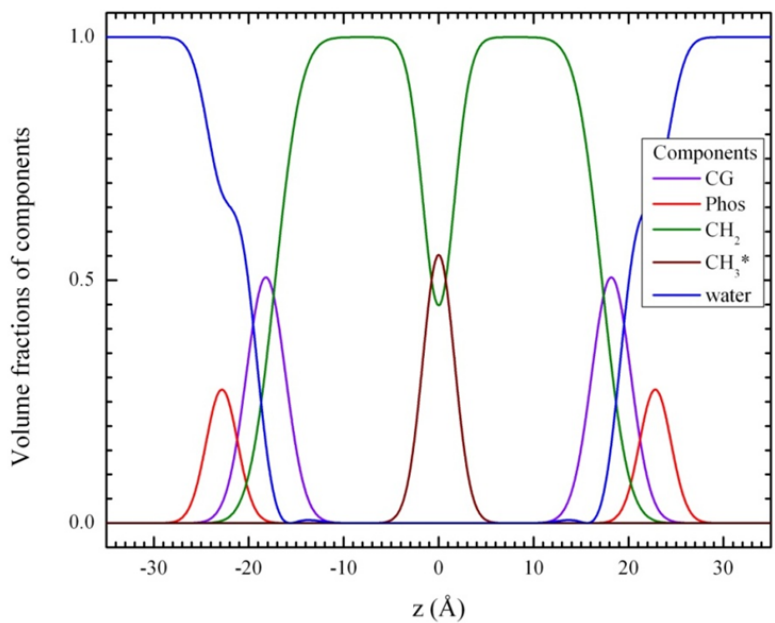

Fig. 5. Component volume fractions vs distance z from bilayer center.

An historically important quantity is the ratio $r$ of methyl volume to methylene volume. The conventional way of parsing the hydrocarbon chain results in $r = 2.08$. The new parsing gives the same value because the pseudo $CH_3^*$ volume has 1.08 times the methylene volume.

Figs. 4 and 5 do not show the choline component. While the other components were robustly determined, the choline component was not. Indeed, constraining the choline width and position to different values led to quite small changes in the $\chi^2$ of the fit. This is not surprising as its electron density is close to that of water and that is what the best fit gave in this analysis. (An earlier fitting model[25] assumed *a priori* that the choline could be part of the water component.)

The SDP fitting program enforces conservation of volume by assigning water to fill the space not occupied by the lipid components. Because the functional forms for the components are constrained to be Gaussians or error functions, negative water may be assigned even though the W term in Eq. (3) penalizes that. This may be attributed to the CG component which assumes the same electron density throughout its volume but which will be more electron dense near the lower carbonyl end than near the upper glycerol end. This suggested dividing the CG group into separate carbonyl and glycerol components which solved the negative water problem, but the positions and widths of the headgroup peaks were contradictory to stereochemistry, which is attributable to having too many overlapping groups in the interfacial region. Instead, the fits in Figs. 3, 4 and 5 were obtained by increasing the width SC of the hydrocarbon interface until the water component was positive for all z.

Table 1 collects the structural parameters for gel phase DPPC at T= 20 °C obtained by fitting the LAXS data.  The volumes are $V_L$ for gel phase DPPC, $V_H$ for the headgroup, $V_{Ph}$ for the phosphate component, $V_{CG}$ for the carbonyl/glycerol component, and $V_{CH2}$ for the average chain methylene volume. The ratio of the chain terminal methyl volume to $V_{VH2}$ is $r$. The area per lipid is A. The Luzzati thickness is $D_B$, the peak in the electron density profile is at $D_{HH}/2$, the phosphate is centered at $z_{Ph}$ with width $\sigma_{Ph}$ and the carbonyl/glycerol component is centered at $z_{CG}$ with width $\sigma_{CG}$. The hydrocarbon interface is centered at $D_C$ and its width is $\sigma_{HC}$ and the width of the hydrocarbon chain terminal methyls is $\sigma_{CH3}$. The number of water molecules per lipid $n_W$ is given for the fully hydrated spacing D = 63.2 Å. For convenience in discussion, the distance between $D_{HH}/2$ and $D_C$ is $D_{H1}$ and two other differences are also listed at the bottom of Table 1.

The second column in Table 1 (labeled $V_H D_C$) lists our preferred values for the quantities in the first column. The asterisks indicate quantities whose values were constrained from volume and wide angle x-ray measurements, and by the above-mentioned non-negative water constraint on $\sigma_{HC}$. Their uncertainties are indicated in the third column. The third column also gives an uncertainty for each unconstrained quantity. That uncertainty was obtained by fixing the value of the quantity different by $\delta$ from its best fitted value and then fitting all the other quantities. Trial and error found the value of $\delta$ for which the total $\chi^2$ was one greater than the smallest $\chi^2$ and that is the uncertainty for that quantity that is listed in column 3. These uncertainties are quite small.  Larger uncertainties ensue by varying the constrained parameters.  Column 4 primarily changes $V_H$ to a previously reported value,[7] and this increases $\chi^2$ somewhat while leaving most other quantities nearly the same within the uncertainties. Column 5 shortens $D_C$ as might occur if the hydrocarbon chains are not in their usually assumed all-trans chain conformation. This hardly affects $D_{HH}/2$ as the electron density profile is quite robust. However, it makes the bilayer thinner and therefore requires it to have a larger area. Column 6 explores increasing $V_H$ to be larger than indicated by wide angle x-ray scattering. While this achieves very slightly smaller $\chi^2$, it hardly changes the other results when compared to the $V_HD_C$ model in column 2. The values of $\sigma_{HC}$ in Table 1 are consistent with a value of $\sigma_{HC} = 2.15$ Å that corresponds to an offset of two carbons (1.27 Å/carbon) along inequivalent all-trans *sn-1* and *sn-2* chains tilted by 32°. Also, the headgroup component volumes agree satisfactorily with the values $V_{Ph} = 54$ Å$^3$ and $V_{CG} = 147$ Å$^3$ obtained from simulations of fluid phases.[28, 29]

| 1 | 2 | 3 | 4 | 5 | 6 |
|---|---|---|---|---|---|
|  | $V_HD_C$ | δ | $V_H2D_C$ | $V_HD_C2$ | $V_H3D_C$ |
| $\chi^2_{reduced}$ | 6.699 |  | 6.815 | 6.713 | 6.692 |
| $V_L$ | 1144* | 1# | 1144* | 1144* | 1144* |
| $V_H$ | 331* | 12# | 319* | 331* | 334* |
| $D_C$ | 17.2* | 1# | 17.4* | 16.2* | 17.2* |
| A | 47.3 |  | 47.4 | 50.2 | 47.1 |
| $D_B$ | 48.4 |  | 48.3 | 45.6 | 48.6 |
| $\sigma_{HC}$ | 2.18* |  | 2.27* | 1.78* | 2.25* |
| $D_{HH}/2$ | 22.63 | 0.03 | 22.62 | 22.65 | 22.61 |
| $z_{Ph}$ | 22.82 | 0.02 | 22.84 | 22.83 | 22.81 |
| $\sigma_{Ph}$ | 1.66 | 0.02 | 1.63 | 1.65 | 1.67 |
| $z_{CG}$ | 18.18 | 0.03 | 18.21 | 18.23 | 18.19 |
| $\sigma_{CG}$ | 2.06 | 0.03 | 2.1 | 2.08 | 2.05 |
| $V_{Ph}$ | 54 | 1.2 | 49.7 | 53.5 | 55.1 |
| $V_{CG}$ | 124 | 2 | 115 | 124 | 126 |
| $V_{CH2}$ | 25.3 | 0.3 | 25.6 | 25.3 | 25.2 |
| r | 2.08 | 0.01 | 2.16 | 2.1 | 2.06 |
| $\sigma_{CH3}$ | 1.67 | 0.01 | 1.68 | 1.68 | 1.67 |
| $n_W$ | 11.7 |  | 11.8 | 14.7 | 11.5 |
| $D_{H1}$ | 5.43 | 0.03 | 5.42 | 6.45 | 5.41 |
| $z_{Ph}- z_{CG}$ | 4.64 | 0.02 | 4.63 | 4.60 | 4.64 |
| $Z_{CG}- D_C$ | 0.98 | 0.03 | 0.81 | 2.03 | 0.97 |

Table 1. Values obtained from fitting models named in the top row to the experimental data. Quantities marked by * were constrained. Uncertainties in δ marked by # are estimated from volume and WAXS measurements. All units are appropriate powers of Å.

## IV. Discussion

The primary new results in this paper are the form factors F(q) shown in Fig. 3. In addition to providing more q values than previously, these data show that there are significant data out to 1.3 Å$^{-1}$, extending beyond the previous q range that ended near 1.0 Å$^{-1}$. These new data provide higher spatial resolution in the electron density profiles in Fig. 4. However, while those profiles represent the Fourier transforms of the data, they were obtained using models. The real data that simulators should compare to are the |F(q)| absolute values, relative to the K$_e$ scale factors for each data set corresponding to different exposures. For convenience, numerical tables are provided in Supplementary Material.

There are also other important experimental results that simulations should compare to. Molecular volume and some of the WAXS results are listed in Appendix A. One additional major WAXS result is that the chains in each monolayer are parallel to those in the opposing monolayer.[7] Another is that the chains are tilted toward nearest neighbors as in the L$_{\beta I}$ structure.[30] Another is that headgroup ordering must be weak due to a lack of the appropriate WAXS scattering.[7] Another is that there is correlation in chain orientation over a distance of 2600 Å.[7] Also hitherto not mentioned is a result for the order parameter g for the azimuthal chain orientation relative to the tilt direction that has been extracted from oriented infrared absorption experiments.[31] The only simulation that has compared to the azimuthal chain direction obtained a different sign for g, although that simulation obtained notably good agreement with |F(q)| for the older range of q and also with the chain tilt angle,[10] so perhaps the experimental value of g should also be re-examined.

If a simulation can obtain satisfactory results for the existing data, and there is a lot of it, then many additional quantities of physical chemical interest that are not available from experiment could be addressed. How wide is the distribution of tilt angles? How spatially disordered is the chain packing unit cell – are there usually six nearest neighbor chains and how much disorder is due to inequivalent *sn-1* and *sn-2* packing? How many gauche rotamers are there? Are there gtg kinks that could shorten the chains, and if so, how are they correlated with the tilt direction? Is there a correlation between chains on the same molecule and also on the tilt direction? Are there azimuthal correlations between nearest neighbor chains? Is there mini-interdigitation (*vide infra*) of the terminal methyl ends of the *sn-1* and *sn-2* chains in the center of the bilayer as suggested by the modeling in this paper? Moving from the chains to the headgroups, is there some weak P-N short range orientational correlation? Is there some P-N orientational correlation with the chains on the same molecule? What is the distribution of the internal glycerol backbone angles and how do they compare to the fluid phase? How far is the phosphate from the hydrocarbon interface ($D_{H1}$ in Table 1)?

One might be pessimistic about simulations ever being able to agree with the existing data when the simulation is initialized from some generic starting point because the force fields are unlikely to be good enough to match all the experimental data, and even if the force fields are perfect, the equilibration time could be very long. Instead, following Venable et al.[8] it seems more promising to deliberately initialize models conforming to the experimental results in the second paragraph above and perform short, essentially local energy minimizations to find a model that comes closest to the new |F(q)| experimental data. To the extent that the obtained model has higher free energy than longer simulations that do not conform to the experimental data would then provide a measure of the inaccuracy of the force field.

This paper has also taken the non-simulation modeling approach to obtaining more detailed information about the DPPC gel phase. One notable result addresses the chain packing in the center of the bilayer. Typical cartoons of gel phase lipid bilayers show all the terminal methyl ends of the hydrocarbon chains ending at the same z level with a total 100% gap between the two opposing monolayers. In contrast, Fig. 5 shows that the methylene and the methylene part of the terminal methyl occupy 50% of the space in the center of the bilayer. Figure 6 shows how mini-interdigitation of the hydrocarbon chains can account for this. In Fig. 6 all-trans four hydrocarbon chains are represented by strings of circles in straight lines. Interior circles represent methylenes and circles at the chain ends represent the methylene part of the terminal methyls. We do not know the azimuthal orientation of the all-trans chains, so Fig. 6 simply shows the orientation that has the zig-zag chain oriented perpendicular to the plane of the figure. The distance between the centers of the circles is 1.27 Å which is the projected length along the all-trans axis of the C-C bonds of length 1.54 Å. The tilt angle and the distance between chains in the same monolayer come from WAXS scattering results.[6,7] Neutron diffraction results gave the z level of the C2 carbons on the *sn-2* chain 1.9 Å further from the center of the bilayer than the C2 carbons on the *sn-1* chain,[2] so the terminal methyls of the all-trans chains are shown to be equally offset in the z-direction. Mini-interdigitation comes about by opposing the sn-1 chains on one monolayer with the sn-2 chains on the other monolayer. There is a gap between the chain ends that is indicated by a double arrow in Fig. 6. The length of that gap is drawn to be the diameter of two circles because each terminal methyl occupies nearly twice the volume of the methylenes, namely the *r* value in Table 1. With these quantities established, Fig. 6 shows that the methylene probability density is half as large in the $|z| < 2.1$ Å region between the red dashed lines in Fig. 6 as it is $|z| > 2.1$ Å, in good agreement with Fig. 5. As discussed earlier,[6] miniinter-digitation would provide a structural linkage between the two monolayers that might account for the WAXS result that the chains in opposing monolayers are tilted in the same direction.[7] This is the strongest evidence we have found for mini-interdigitation in DPPC gel phase. It may also be recalled that the lipid MPPC doesn't even have a gel phase;[32] this is consistent with MPPC having two fewer methylenes on its *sn-1* chain than on its *sn-2* chain, and therefore likely to have less mini-interdigitation.

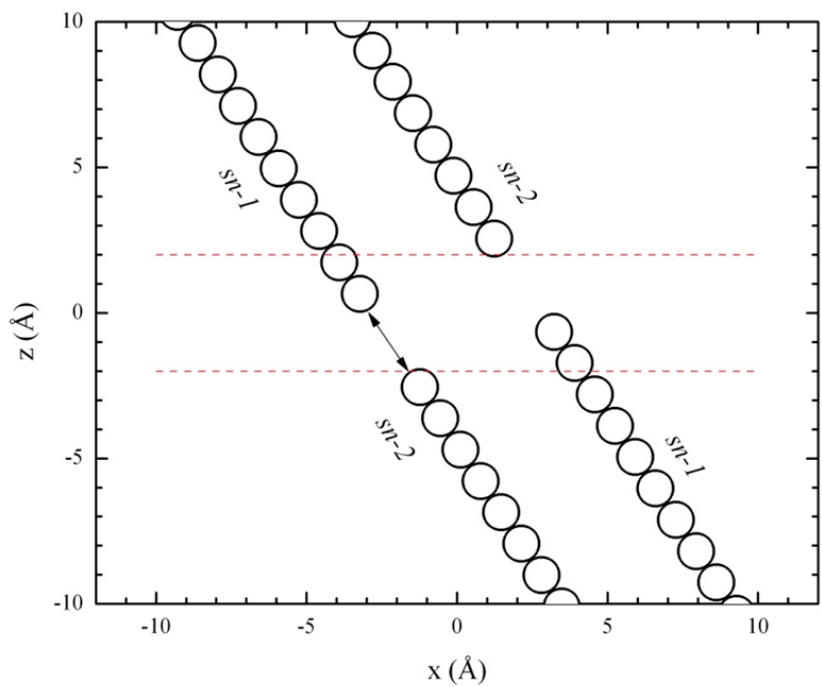

Figure 6. Illustration of mini-interdigitated chain packing in the center of the bilayer delineated by the dashed lines.

Another notable modeling result is that the greater range of |F(q)| data gives higher resolution in the electron density profile in Fig. 4. In particular, the two headgroup peaks due to the phosphate and the carbonyl/glycerol components are now more sharply defined and better separated than in an earlier DPPC profile[4] and in the best DMPC profile.[20] This means that $D_{HH}/2$ is less influenced by the CG electron density. One of the consequences is that the $D_{H1} = D_{HH}/2 - D_C$ is now larger for DPPC than the value 5.0 Å previously obtained for DPPC[4] or than the 4.95 Å value obtained for gel phase DMPC.[20] The nice agreement in the previous values for DPPC and DMPC suggested that $D_{H1}$ was a robust quantity that could be used to constrain analysis of fluid phases of PC lipids to obtain area per lipid A.[21] However, neutron scattering more straightforwardly obtains values of A that require a smaller value of $D_{H1}$, especially for DOPC.[33] Now, this new larger value of $D_{H1}$ for DPPC gel phase also reinforces not using gel phase $D_{H1}$ in future to bootstrap fluid phase structure from gel phase structure. The new result suggests that the headgroup conformation is different in the gel and fluid phases, something that simulations might address. While there is substantial difference in values of $D_{H1}$ between DMPC and DPPC gel phase, it is encouraging that their r values are essentially identical, and they agree very well with values from liquid alkanes.[29] The areas A are very similar when taking into account that DMPC was measured at a lower T = 10°C.[20] The number of water molecules per lipid in fully hydrated multilamellar vesicles also round to $n_W$ = 12 for DMPC and DPPC. These are considerably smaller than for the fluid phase ($n_W$ = 27 for DMPC[21] and 30 for DPPC[34]) due to suppression of the repulsive undulational fluctuation force by the stiffer gel phase bilayers.[35] Overall, the comparison of DMPC and DPPC suggests that gel phases in same chain PC lipids are similar and the current data once again make DPPC gel phase the pre-eminent one.

Acknowledgements: We thank beamline scientist Arthur Woll for facilitating our collecting X-ray scattering data at the G1 station of the Cornell High Energy Synchrotron

Source (CHESS), which is supported by the National Science Foundation and the National Institutes of Health/National Institute of General Medical Sciences under National Science Foundation Award DMR-0225180.

Appendix: Literature results for other quantities used in modeling

Table 2 assembles published results obtained from wide angle scattering and volumetric measurements from the references listed in column 1 at the temperatures in column 2. All rows are for DPPC gel phase except for the DMPC row and the final rows show the values used for the models in the main text. The third column in Table 2 shows the volume of the lipid $V_L$. The fourth column shows the area per chain $A_C$ measured perpendicular to the tilted chains with tilt $\theta_t$ in the 7$^{th}$ column. $A_C$ and $\theta_t$ are the two main WAXS results. For oriented samples $A_C$ was obtained from the $d_{11}$ and $d_{20}$ spacings and $\theta_t$ was obtained from the elevation of the (11) peak from the equator. The volume of the hydrocarbon chains $V_C$ was calculated as $2(14+r)(1.27\text{Å})A_C$; the length of a single all-trans chain consisting of 14 methylenes and a terminal methyl is $(14+r)(1.27\text{Å})$ where $r=2$ accounts for the extra van der Waals length of the terminal methyl. Column 6 shows the volume of the headgroup $V_H = V_L - V_C$. The area per lipid at the interface, $A_L = 2A_C/\cos\theta_t$, is shown in column 8. The last column gives the volume per methylene $V_{CH2} = (1.27\text{Å})A_C$. The italicized numbers were not specifically written in the primary references but can easily be calculated from quantities given in those references.

|  | T (°C) | $V_L$ (Å³) | $A_C$ (Å²) | $V_C$ (Å³) | $V_H$ (Å³) | $\theta_t$ (°) | $A_L$ (Å²) | $V_{CH2}$ (Å³) |
|---|---|---|---|---|---|---|---|---|
| DPPC[26] | 20 | 1144±2 | 19.9±0.2 | *804±10* | 340±10 | 30±3 | 45.9±2.0 | 25.3±0.2 |
| DPPC[6] | 19 | *1144±2* | 20.0±0.1 | *813±4* | *331±6* | 32.0±0.5 | 47.2±0.5 | *25.4* |
| DPPC[7] | 24 | 1148±2 | 20.4±0.04 | 829±4 | 319±6 | 31.6±0.2 | 47.9±0.2 | 25.9±0.1 |
| DPPC[5] | 20 | *1144±2* | 20.1±0.2 | *817±8* | *327±8* | 32.0±0.5 | *47.2* | 25.5 |
|  | 25 | *1148±2* | 20.2±0.2 | *821±8* | *327±8* | 31.6±0.4 | 47.3±0.3 | *25.7* |
| DMPC[20] | 10 | 1041 | *19.9* | *706* | 331 | 32.3 | 47.0 | 25.3 |
|  |  |  | *20.1* | *714* | 319 | 32.3 | 47.5 | 25.5 |
| $V_HD_C$ | 20 | 1144 | 20.1 | 813 | 331 | 32.0 | 47.3 | 25.3 |
| $V_H2D_C$ |  |  | 20.1 | 825 | 319 | 32.0 | 47.4 | 25.6 |
| $V_H3D_C$ |  |  | 20.0 | 810 | 334 | 32.0 | 47.1 | 25.2 |

Table 2. Literature and modeling results.